\documentstyle[12pt]{article}

\begin{document}

\title{On Discrete Quasiprobability Distributions}
\author{C.A. Mu\~noz Villegas${}^1$, A. Chavez Chavez${}^1$, S. Chumakov${}^{1}$, \\
Yu. Fofanov${}^2$, A.B. Klimov${}^{1}$}
\maketitle

\begin{center}
${}^1$Departamento de F\'{\i}sica, Universidad de Guadalajara, M\'exico%
\\[0pt]
${}^2$ Computer Science Department, University of Houston, Texas, USA
\end{center}

\begin{abstract} 
     We analyse quasiprobability distributions related to the discrete Heisenberg-Weyl group. 
     In particular, we discuss the relation between the Discrete Wigner and Q-functons.
\end{abstract}

\section{Introduction}

Due to the recent development of hardware facilities, the description of the
state and evolution of the wave processes has also evolved. In addition to
the configuration space description and Fourier analysis,  it has now become
possible to use the Quasiprobability distributions that describe the state
of the wave process in phase space of the corresponding mechanical system.
By wave process we mean any acoustical, optical or quantum mechanical
process, where, correspondingly, time and frequency, the coordinate of the
point where the ray of light intersects the screen and the ray direction,
the coordinate and momentum of the quantum mechanical particle play the role
the phase space coordinates (see, {\it e.g.} [1-6]).  Until quite recently,
the practical use of the Quasiprobability distributions was strongly limited
by the speed and memory restrictions of the computers, but now the situation
is changing. The possibility to use the Quasiprobability distributions is
highly attractive for it allows us to visualize the process in a intuitively
clear way: to analyze the signal ({\it e.g.} to discriminate between its
different components, possibly originated from different sources) and to
process the signal (say, removing noise) at the level which is much higher
than the one that can be achieved when working with the signal itself or its
Fourier transform.

The theory of Quasiprobability distributions in its most complete form was
developed in the framework of quantum mechanics ({\it cf.}
[1-4,6]). In its original version, it refers to continuous variables and is
intrinsically related to the Heisenberg-Weyl group of translations of phase
space. However, any numerical work involves the discretization, which by
itself is not a trivial procedure, as is clear from the example of the
approximation of the Fourier integral by the Finite Fourier transform (see,
{\it e.g.} \cite{Wolf}). Therefore, it is desirable to develop a systematic
theory of the Quasiprobability distributions in discrete phase spaces. The
basis of such a theory was given in the work of Wootters, \cite{Wootters}
(see also \cite{Leanhart}, \cite{Vourdas1}, \cite{Vourdas2}, \cite{Paz1},
\cite{Paz2}). The Hilbert space of system states was chosen to be a space of
periodic functions with a finite number of Fourier harmonics. This is
obviously a finite-dimensional space.  This type of quasiprobability
distributions is naturally related to the discrete Heisenberg-Weyl group of
translations of discrete phase space. (It should not be confused with the
quasiprobability distributions for spin systems, which are related to the
SU(2) group \cite{Stratonovich,Klimov}; in this case, the finite-dimensional
Hilbert space includes spherical harmonics on the sphere.)

Our goal in this work is to establish the correspondence between the two
most important discrete Quasiprobability distributions: the Wigner-Wootters
(``W-function'') distribution and the time dependent spectrum
(``Q-function''), as it exists for the continuous case, where the Q-function
can be produced by smoothing the W-function with an appropriate Gaussian. A
natural periodic counterpart of the Gaussian is the Jacobi Theta-function
and therefore, they appear in our construction.


The paper is organized as follows. After the presentation of the discrete
coherent states (which naturally involves the discrete Jacobi
Theta-functions) and the Q-function, we define the discrete Wigner function
(which is made in terms of the discrete Heisenberg-Wyel group and is
complementary to the original work \cite{Wootters}). Finally, we establish
the relation between them.


We would like to note that we do have in mind possible applications to
signal processing. However, at this stage of theoretical development, we
found it reasonable to confine ourselves within the language of Quantum
mechanics.

\section{The model}

We consider periodic functions $f(x + L) = f(x).$ Therefore, the Fourier
series coefficients contain the frequencies, $\omega_k = k 2\pi/L$, where $k$
is an integer,
\[
f(x)= \sum_k \underline{f}_{\,k} e^{i x 2\pi k/L}, \quad \underline{f}_{\,k}
= \frac{1}{L}\, \int_0^L dx \,f(x) e^{-i x 2\pi k/L}.
\]

Let us consider functions for which Fourier series contain only $M$
coefficients different from zero: $\; k=0, \, 1, \ldots, M-1. $ Then all the
information is stored in the values of the function in $M$ points on the
circle:
\[
x_m = \frac{L m}{M}, \quad m=0,1,\ldots, M-1;
\]
\[
f(m)\equiv f(x_m)= \sum_{k=0}^{M-1} \underline{f}_{\,k}  e^{i 2\pi k m/M} .
\]
The values of the function $f(x)$ at arbitrary points can be recovered by
the sampling theorem.

We will use the quantum mechanics notation, introducing the basis $|{m_0}%
\rangle=\delta_{mm_0}$,
\[
|{f}\rangle = \sum_{m=0}^{M-1} |{m}\rangle \langle m|{f}\rangle, \quad
\langle m|{f}\rangle=f(x_m)=f(m).
\]
One can use the matrix representation with
\[
|{f}\rangle = \left|
\begin{array}{c}
f_0 \\
f_1 \\
\ldots \\
f_{M-1}
\end{array}
\right|.
\]

The Discrete Fourier Transform of the vector $f(m)$ gives us the wave
function in momentum representation:
\[
f(p)= \langle p|{f}\rangle = \sum_{m=0}^{M-1}  \langle p|{m}\rangle \langle
m|{f}\rangle, \quad  p=0,1,\ldots, M-1, \quad  \langle p|{m}\rangle = \frac{1%
}{\sqrt M} e^{-i2\pi p m/M}.
\]

Using the discrete orthogonality relation (\ref{DiscOrt}),

\begin{equation}  \label{DiscOrt}
\sum_{m=0}^{M-1} e^{i2\pi m(n-k)/M} = M\delta_{n,k (mod\, M)},
\end{equation}
one finds the DFT coefficients,

\[
f(p)= \sqrt{M} \underline{f}_{\,p\, (mod \,M)}
\]
Thus, the periodicity also appears in the momentum.

It is natural to introduce one-step translations in coordinate,
\[
\hat T_x f(x_m) = f(x_{m-1}), \quad \hat T_x = \exp \left(-i\,\frac{L}{M}%
\,\hat p_x\right), \quad \hat p_x = -i\partial_x , \quad (\hbar = 1),
\]
and in momentum,
\[
\hat T_p = \exp \left(i\,\frac{2\pi}{L}\,\hat x\right), \quad \hat x \, f(x)
= x f(x) , \quad \hat T_p f(x_m) = r^{-m} f(x_{m}), \quad r=e^{i 2\pi/M}.
\]
It is clear, that
\[
\hat T_p \hat T_x = \sqrt{r}\, \hat T(1,1) = r \hat T_x \hat T_p , \quad
\hat {\underline{T}}(1,1) = \exp \left[ i\left( \frac{2\pi}{L}\,\hat x -
\frac{L}{M}\,\hat p_x \right)\right].
\]
Note, that in the course of the discretization of the initial continuous
model, $L$ completely disappears from all of the formulas.

\section{Vacuum state}

The periodic analogue to Gaussian functions is the Jacobi $\Theta$-function
defined as follows (see \cite{WhWat} and Appendix A),

\[
\Theta_3(z,\mu)  = \sum_{n=-\infty}^{\infty} e^{-i2z n - \mu n^2}  = \sqrt{%
\frac{\pi}{\mu}}\, \sum_{k=-\infty}^{\infty}  \exp\left\{-\frac{(z - \pi k)^2%
}{\mu}\right\}.
\]
(Here and below we often use the Poisson transformation.) Generally
speaking, Re$\,\mu>0$ but we consider only real values of $\mu$. Recall,
that $\Theta_3(z+\pi,\mu)=\Theta_3(z,\mu)$, and hence $x=z L/\pi$.

Considering the $\Theta$-function on a discrete set of points, $x_m=mL/M$
and $z_m=m\pi/M$, we come to a discrete periodic Gaussian,
\begin{eqnarray}
\Theta(m) &=& \langle{m}|{\Theta_{\mu}}\rangle = \Theta_3\left(\left.\frac{%
\pi m}{M}\right|\mu\right)= \sum_{n=-\infty}^{\infty} e^{-i2\pi m n/M - \mu
n^2}  \nonumber \\
&=& \sqrt{\frac{\pi}{\mu}}\, \sum_{k=-\infty}^{\infty} \exp\left\{ - \frac{%
\pi^2(m - kM)^2}{M^2\mu}\right\}.  \label{OCS}
\end{eqnarray}
This function will play a role of the vacuum squeezed state (with squeezing
parameter $\mu$). The choice
\[
\mu = \pi/M
\]
corresponds to the vacuum Coherent State. Thus, $\mu\ll 1$, $\mu M^2 \gg 1$
will be of interest. Using the discrete orthogonality relation (\ref{DiscOrt}%
) one can check the normalization,
\begin{eqnarray}
{\cal N} &=& \langle \Theta_{\mu}|{\Theta_{\mu}}\rangle = \sum_{m=0}^{M-1}
|\Theta(\phi_m|\mu)|^2 = M\sum_n\sum_k e^{-\mu n^2 -\mu (n-kM)^2}  \nonumber
\\
&=& M \left\{
\begin{array}{ll}
\left[\theta_3(2\mu) \theta_3(2\mu M^2) + \theta_2(2\mu) \theta_2(2\mu M^2)
\right], & \; M=\hbox{odd} \\
\theta_3(2\mu) \theta_3(\mu M^2/2), & \; M=\hbox{even}
\end{array}
\right.  \nonumber \\
&=& \sqrt{\frac{2\pi}{\mu}} \left\{
\begin{array}{ll}
\left[\theta_3(2\mu) \theta_3(\mu^{\prime}/2) + \theta_2(2\mu)
\theta_4(\mu^{\prime}/2) \right]/2 , & \; M=\hbox{odd} \\
\theta_3(2\mu) \theta_3(2\mu^{\prime}), & \; M=\hbox{even}
\end{array}
\right.  \label{CSnorm}
\end{eqnarray}
Here
\[
\mu^{\prime}= \frac{\pi^2}{\mu M^2} ;
\]
for unsqueezed vacuum CS, $\mu=\mu^{\prime}=\pi/M.$ We will assume, that $%
\mu = \mu_0 \pi/M,$ $\mu_0\sim 1$, and thus, $\mu,\mu^{\prime}\ll 1$
simultaneously with $\mu M^2 , \mu^{\prime}M^2 \gg 1$.

In Eq.\ (\ref{CSnorm}) we use the notation,


\[
\theta_k(\mu) = \Theta_k(0|\mu), \quad k =1,2,3,4;
\]
(see Appendix A). For instance,
\[
\theta_3(\mu)= \sum e^{-\mu n^2}  = \sqrt{\frac{\pi}{\mu}}\sum e^{-
k^2\pi^2/\mu}  = \sqrt{\frac{\pi}{\mu}} + O\left( e^{- \pi^2/\mu}\right) ,
\]
\[
\theta_2(\mu)= \sum e^{-\mu (n+1/2)^2}  = \sqrt{\frac{\pi}{\mu}}\sum (-1)^k
e^{- k^2\pi^2/\mu}  = \sqrt{\frac{\pi}{\mu}} + O\left( e^{-
\pi^2/\mu}\right) .
\]
In passing from the second to the third line of Eq.\ (\ref{CSnorm}) we again
used the Poisson transformation, $\theta_3(\mu M^2/2)=\sqrt{2\mu^{\prime}/\pi%
}\,\theta_3(2\mu^{\prime})$, and $\theta_2(2\mu M^2)=\sqrt{\mu^{\prime}/2\pi}%
\,\theta_4(\mu^{\prime}/2)$. If $M\gg1$, the normalization constant has a
simple asymptotic form. Indeed, in this case only the terms $k=0$ are
important in the Poisson-transformed expressions for $\theta_{2,3,4}$,
and neglecting the terms $O\left(e^{-\mu M^2/2}\right)$, $%
O\left(e^{-\mu^{\prime}M^2/2}\right)$, we have,

\begin{eqnarray}  \label{CSnorm1}
\langle \Theta_{\mu}|{\Theta_{\mu}}\rangle \approx M\sqrt{\frac{\pi}{2\mu}}.
\end{eqnarray}

The wave function of the vacuum state in the momentum representation is
given by the Discrete Fourier transform:
\[
\langle p|{\Theta_{\mu}}\rangle  = \sqrt{M} \sum_{k=-\infty}^{\infty}
e^{-\mu (p- kM)^2}  = \sqrt{\frac{\pi}{\mu M}} \sum_{n=-\infty}^{\infty}
e^{-i2\pi pn/M - \mu^{\prime}n^2}  = \sqrt{\frac{\pi}{\mu M}} \,
\Theta\left( \left. \frac{\pi p}{M}\right| \mu^{\prime}\right).
\]


\section{Discrete Heisenberg-Weyl group}

Let us return to the one-step translations in coordinate and momentum, which
are defined in the framework of the initial continuous model as $\hat T_x =
e^{-i \hat p_x L/M}$, $\hat T_p = e^{i\hat x 2\pi/L}$,
\[
\hat T_x|{m}\rangle=|{m+1}\rangle,\quad \hat T_x f(m)=\langle{m}| \hat T_x |{%
f}\rangle =f(m-1),\quad \hat T_x|{p}\rangle = r^{-p}|{p}\rangle,\quad
\]
\[
\hat T_p|{m}\rangle=r^m|{m}\rangle,\quad
\hat T_p|{p}\rangle = |{p+1}\rangle,\quad \hat T_p f(p)=f(p-1).
\]
Here, once again, $r=e^{i 2\pi/M}$. These operators generate a discrete
group. It is clear that there is a periodicity in momentum, as well as in
coordinate, $\hat T_p^M =\hat T_x^M =1$. These translations do not commute,
\[
\hat T_p^n\hat T_x^m = r^{nm} \hat T_x^m\hat T_p^n.
\]
Therefore, an arbitrary group element must include the multiplication by a
phase factor,
\begin{equation}  \label{gHW}
g(s,m,n)= r^{s} \hat T_x^m\hat T_p^n.
\end{equation}

The naive way to introduce the displacement operator in phase space is
(compare \cite{Vourdas1}, \cite{Vourdas2}),
\[
\hat{\underline{T}}(m,n)=r^{nm/2}\hat{T}_{x}^{m}\hat{T}_{p}^{n}=r^{-nm/2}%
\hat{T}_{p}^{n}\hat{T}_{x}^{m}=\exp \left[ i\left( \frac{2\pi }{L}\hat{x}n-%
\frac{L}{M}\hat{p}m\right) \right] .
\]
However, such an operator does not belong to the group (\ref{gHW}).
Fortunately, there exists a way to improve the situation. Let us now
consider the case of odd $M$. Then, in the set of numbers $%
\{n\,(modM)\}=\{1,2,\ldots ,M-1\}$ there exists a unique solution to the
{\bf equation} $2n=m$, which we denote as
\[
\left[ \frac{m}{2}\right] =
\begin{array}{ll}
\frac{m}{2} & m=2k, \\
\frac{m+M}{2} & m=2k+1.
\end{array}
\]
(More generally, in the case of prime $M$ there always exists a unique
solution to the equation $sn=m$, and the numbers $\{n\,(modM)\}$ then form a
field.) It is clear, that $\left[ \frac{mn}{2}\right] =\left[ \frac{m}{2}%
\right] n=\left[ \frac{n}{2}\right] m$ and $\left[ \frac{m+n}{2}\right]
=\left[ \frac{m}{2}\right] +\left[ \frac{n}{2}\right] $.

We now introduce the displacement operator as follows,
\begin{equation}  \label{HW}
\hat T(m,n) =r^{[nm/2]} \hat T_x^m\hat T_p^n = r^{-[nm/2]} \hat T_p^n\hat
T_x^m. =(-1)^{mn} \hat{\underline{T}}(m,n).
\end{equation}
$\hat T$ is unitary, $\hat T^{\dagger}(m,n)=\hat T^{-1}(m,n)=\hat T(-m,-n)$,
and periodic, $\hat T(m+M,n)=\hat T(m,n+M)=\hat T(m,n)$. It is easy to check
the multiplication formula,
\begin{equation}  \label{HW1}
\hat T(m_2,n_2) \hat T(m_1,n_1) = r^{[(m_1n_2-n_1m_2)/2]} \hat
T(m_1+m_2,n_1+n_2).
\end{equation}
One can introduce the adjoint action,

\begin{eqnarray}  \label{Ad}
\hat T(q,p)\hat T(m,n)\hat T^{\dagger}(q,p) = r^{p m-q n } \hat T(m,n).
\end{eqnarray}
The matrix elements in the coordinate basis,
\[
\langle{k}| \hat T(m,n)|{l}\rangle = r^{nl+[nm/2]} \delta_{k,l+m (mod M)}.
\]
These matrix elements are orthogonal,
\begin{eqnarray}  \label{MEOrt}
\sum_{m,n=0}^{M-1} \langle{a}| \hat T(m,n)|{b}\rangle \overline{\langle{d}|
\hat T(m,n)|{c}\rangle} = M\delta_{ad}\delta_{bc}.
\end{eqnarray}
Finally,
\[
\hbox{Tr}\, \hat T(m,n) = M\delta_{m,0}\delta_{n,0}.
\]

\section{Discrete CS and Q-dunction}

One can generate the complete set of (squeezed) CS by the action of the HW
group (\ref{HW}) to the vacuum state (\ref{OCS}),
\[
|{m_0,n_0,\mu}\rangle = \hat T(m_0,n_0) |{\Theta_{\mu}}\rangle,
\]
\begin{eqnarray}  \label{CS}
\Theta_{m_0n_0}(m) = \langle m |{m_0,n_0,\mu}\rangle = e^{i 2\pi n_0 (m -
m_0/2)/M } \, \Theta(m-m_0) \\
= e^{i 2\pi n_0 (m - m_0/2)/M } \, \sum_{n = - \infty}^{\infty} \exp \left[
i\, \frac{2\pi}{M}\,(m_0-m) n -\mu n^2 \right],
\end{eqnarray}
and in the momentum representation,
\begin{eqnarray}
\Theta_{m_0n_0}(p) = \langle p |{m_0,n_0,\mu}\rangle =\sqrt{M} \, e^{-i 2\pi
(p - n_0/2)m_0/M }\, \sum_{k=-\infty}^{\infty} \exp \left[-\mu
(p-n_0-kM)^2\right] .  \nonumber
\end{eqnarray}
The normalization constant $\langle{m_0,n_0,\mu} |{m_0,n_0,\mu}\rangle =%
{\cal N} $ is the same as in (\ref{CSnorm}).

\vspace{2mm}

{\it The completeness relation} holds (for any $\mu$):

\begin{equation}  \label{CR}
\frac{1}{{\cal N}}\, \sum_{m_0,n_0=0}^{M-1} |{m_0,n_0,\mu}\rangle \langle{%
m_0,n_0,\mu}| = \hat 1.
\end{equation}

\vspace{2mm}

{\it Scalar products of Coherent States.} \quad It is enough to consider the
vacuum matrix element of the displacement operator,
\[
\langle \Theta_{\mu}|{m_0,n_0,\mu}\rangle  = \langle{\ \Theta_{\mu}}| \hat
T(m_0,n_0) |{\Theta_{\mu}}\rangle .
\]
If $M$ is {\it even}, it is equal to
\[
\langle \Theta_{\mu}|{m_0,n_0,\mu}\rangle=  M\sqrt{\frac{2\mu^{\prime}}{\pi}}
\,\Theta_{\alpha}\left(\left.\frac{\pi m_0}{M} \right| \,2\mu\right)
\,\Theta_{\beta}\left(\left.\frac{\pi n_0}{M} \right| \,2\mu^{\prime}\right)
,
\]
where,
\[
\left\{
\begin{array}{ll}
\alpha=3, & n_0 -\hbox{even}; \\
\alpha=2, & n_0 -\hbox{odd} ;
\end{array}
\right.  \quad \left\{
\begin{array}{ll}
\beta=3, & m_0 -\hbox{even}; \\
\beta=2, & m_0 -\hbox{odd} .
\end{array}
\right.
\]
If $M$ is {\it odd}, $n_0$ is {\it even},
\[
\langle \Theta_{\mu}|{m_0,n_0,\mu}\rangle=  M\sqrt{\frac{\mu^{\prime}}{2\pi}}
\left[  \,\Theta_{3}\left(\left.\frac{\pi m_0}{M} \right| \,2\mu\right)
\,\Theta_{3}\left(\left.\frac{\pi n_0}{2M} \right| \frac{\mu^{\prime}}{2}%
\right)  + (-1)^{m_0}  \,\Theta_{2}\left(\left.\frac{\pi m_0}{M} \right|
\,2\mu\right)  \,\Theta_{4}\left(\left.\frac{\pi n_0}{2M} \right| \frac{%
\mu^{\prime}}{2} \right) \right],
\]
and if $M$ is {\it odd}, $n_0$ is {\it odd},
\[
\langle \Theta_{\mu}|{m_0,n_0,\mu}\rangle=  M\sqrt{\frac{\mu^{\prime}}{2\pi}}
\left[ (-1)^{m_0}  \,\Theta_{3}\left(\left.\frac{\pi m_0}{M} \right|
\,2\mu\right)  \,\Theta_{4}\left(\left.\frac{\pi n_0}{2M} \right| \frac{%
\mu^{\prime}}{2}\right)  + \,\Theta_{2}\left(\left.\frac{\pi m_0}{M} \right|
\,2\mu\right)  \,\Theta_{3}\left(\left.\frac{\pi n_0}{2M} \right| \frac{%
\mu^{\prime}}{2} \right) \right],
\]
Asymptotic for large $M$ is
\[
\frac{\langle \Theta_{\mu}|{m_0,n_0,\mu}\rangle }{ \langle \Theta_{\mu}|{%
\Theta_{\mu}}\rangle} \;  \approx \; \exp\left\{ -\frac{\mu^{\prime}m_0^2+%
\mu n_0^2}{2} \right\} .
\]

\vspace{2mm}

%
%

{\it Q-function.} \quad It is natural to introduce the Q-function as the
diagonal matrix element of the density matrix between the coherent states,
\begin{equation}  \label{QO}
Q(m,p) = \langle m,p,\mu|\rho |{m,p,\mu}\rangle,
\end{equation}
or, for a pure state, $\rho=|{\Psi}\rangle\langle{\Psi}|$,
\[
Q(m,p) = \left|\langle \Psi|{m,p,\mu}\rangle\right|^2.
\]


\section{Wigner function}

By analogy with the continuous case, let us introduce the Wigner operator as
a two-dimensional discrete Fourier transform of the displacement operator,
\begin{equation}  \label{WO}
\hat W(q,p) = \frac{1}{M}\, \sum_{m,n=0}^{M-1} r^{pm-qn} \, \hat T(m,n),
\quad q,p = 0,1,\ldots , M-1.
\end{equation}
Therefore, it is an Hermitian operator valued function on the phase space, $%
\hat W(q,p) = \hat W^{\dagger}(q,p)$. One can notice that,
\[
\hat W(q,p) = \hat T(q,p)\hat W(0,0) \hat T^{\dagger}(q,p).
\]
One immediately calculates its matrix elements,
\[
\langle{k}|\hat W(q,p) |{l}\rangle = \exp\left\{ \frac{i2\pi}{M}%
\,p(k-l)\right\}  \,\delta_{2q=k+l (mod M)}.
\]
These are precisely the matrix elements of the "Phase Point Operators"
considered by Wootters \cite{Wootters}, with the property
\begin{equation}  \label{PP}
\hbox{Tr}\, \left(\hat W(q,p) \hat W(p_1,q_1)\right) = M \delta_{q,q_1}
\delta_{p,p_1}.
\end{equation}
This property was used to define these operators in \cite{Wootters}.

The Wigner function (quasiprobability distribution) on the discrete phase
space for the state $|{f}\rangle$ is the mean value of the Wigner operator
in this state,
\[
W_f(p,q) = \langle{f}|\hat W(q,p) |{f}\rangle.
\]
To an arbitrary Hermitian operator $\hat A$, corresponds the Wigner function
\[
W_f(p,q) = \hbox{Tr} \left(\hat W(q,p) \hat A\right) .
\]
If $\hat A\rightarrow \rho $ is a density matrix and \ Tr $\rho=1,$ then
\[
\sum_{p,q}\hat W(q,p) = 1.
\]
For a pure state, \ Tr $\rho^2=1$,
\[
\sum_{p,q}\hat W^2(q,p) = 1.
\]
The orthogonality of the Wigner operator at different phase points (\ref{PP}%
), has a simple physical sense: the Wigner function of the Wigner operator
itself is a $\delta$-function,
\[
W_{\hat W(p_0,q_0)} (p,q) =M\delta_{q,q_0} \delta_{p,p_0}
\]

{\it Covariance under HW group. } Using the adjoint action of the discrete
HW group, (\ref{Ad}), one can show that
\[
\hat W(q,p) = \hat T(p,q) W(0,0) \hat T^{\dagger}(p,q) .
\]
The Wigner operator at the origin, $W(0,0)$ has a simple form, which we will
explain, considering examples of odd and even $M$. For $M=5$,
\[
W(0,0) = \left|
\begin{array}{ccccc}
1 & 0 & 0 & 0 & 0 \\
0 & 0 & 0 & 0 & 1 \\
0 & 0 & 0 & 1 & 0 \\
0 & 0 & 1 & 0 & 0 \\
0 & 1 & 0 & 0 & 0
\end{array}
\right| ,
\]
thus the spectrum of Wigner operator for odd $M$ consists of $M/2$
eigenvalues $-1$ and $M/2+1$ eigenvalues $+1$. For $M=4$,
\[
W(0,0) = \left|
\begin{array}{cccc}
1 & 0 & 0 & 0 \\
0 & 0 & 0 & 1 \\
0 & 0 & 1 & 0 \\
0 & 1 & 0 & 0
\end{array}
\right| ,
\]
the spectrum for even $M$ consists of $(M-1)/2$ eigenvalues $-1$ and $%
(M-1)/2+1$ eigenvalues $+1$. Therefore,
\[
\hbox{Tr}\, \hat W(q,p) = \sum_{m=0}^{M-1} \langle{m}|\hat W(q,p) |{m}%
\rangle =  1,  = \left\{
\begin{array}{cc}
1, & \quad M=\hbox{odd}, \\
2, & \quad M=\hbox{even},
\end{array}
\right. ,
\]
As in the continuous case, $\hat W(0,0)$ is the inversion operator
with respect to the origin. In the case of odd $M$, there is only
one stable point ($m=0$) under this inversion, while for even $M$
there are two stable points $m=0,$ and $m=M/2$. In the rest of the
paper we will restrict ourselves with the case when $M$ is odd.
(In the opposite case the information is lost and one cannot
reconstruct the state from its Wigner function.)

 {\it Overlap
relation. } Let us consider the following operator acting in the
tensor product of two spaces,
\[
\hat \sigma = \frac{1}{M}\, \sum_{p,q} \hat W(q,p) \times \hat W(q,p).
\]
Using the orthogonality of the displacement operator matrix elements (\ref
{MEOrt}), one can prove that $\hat\sigma$ is an exchange operator,
\[
\langle{a}|_1\langle{c}|_2\hat\sigma|{d}\rangle_2|{b}\rangle_1
=\delta_{b,c} \delta_{a,d}, \quad \hat\sigma|{f}\rangle_1|{g}\rangle_2 =|{g}%
\rangle_1|{f}\rangle_2.
\]

From here the overlap relation follows directly. Let $W_A(p,q)$
and $W_B(p,q) $ be the Wigner symbol of the operators $\hat A$ and
$\hat B$. Then
\begin{eqnarray}  \label{Overlap}
\frac{1}{M}\, \sum_{p,q} W_A(p,q) W_B(p,q) = \hbox{Tr}\ \hat A\hat B .
\end{eqnarray}

\vspace{2mm}

{\it Reconstruction of the initial state.} If we know the Wigner function $%
W_{\rho}(p,a)$, then the density matrix of this state can be reconstructed
as
\[
\hat\rho = \sum_{p,q} \hat W(q,p) W_{\rho}(p,q).
\]

\vspace{2mm}

%

%
%
%

\section{Relation between W and Q-functions}

\vspace{1mm}

We assume now that $M$ is odd, and $\mu=\mu'=\pi/M$. Let us start
with the kernel
\[
\langle{k}|\hat Q(q,p)|{l}\rangle=  \langle k|{q,p}\rangle\langle{q,p}%
|l\rangle  =\langle{k}|\hat T(q,p)|{\Theta_\mu}\rangle \langle{\Theta_\mu}|
\hat T(q,p)^{\dagger}|{l}\rangle,
\]
and find its Fourier transform,
\begin{eqnarray*}
\sum\limits_{q,p=0}^{M-1} r^{qn-pm} \langle{k}|\hat Q(q,p)|{l}\rangle = M^2
\langle{k}|\hat T(m,n) |{l}\rangle F(m,n) ,
\end{eqnarray*}
where the function $F(m,n)$ is defined as
\begin{equation}  \label{Ffun}
F(m,n) = \sqrt{\frac{\mu^{\prime}}{2\pi}}\left[ \Theta_3\left(\left. \frac{%
\pi m}{M}\right|2\mu\right) \Theta_3\left(\left. \frac{\pi}{M}\left[\frac{n}{%
2}\right]\right|\frac{\mu^{\prime}}{2}\right) + (-1)^m \Theta_2\left(\left.
\frac{\pi m}{M}\right|2\mu\right) \Theta_4\left(\left. \frac{\pi}{M}\left[%
\frac{n}{2}\right]\right|\frac{\mu^{\prime}}{2}\right) \right].
\end{equation}
In other words,
\begin{equation}  \label{relQW}
\hat Q(q,p) = \sum_{m,m=0}^{M-1} \hat T(m,n) F(m,n) r^{pm-qn} =
\sum_{m,m=-\infty}^{\infty} \hat T(m,n) f(m,n) r^{pm-qn}.
\end{equation}

\vspace{1mm}

To find the function $f(m,n)$, let us note that for any periodic discrete
function $\phi(m+M)=\phi(m)$ the following formulas hold:
\[
\sum\limits_{n=-\infty}^{\infty} \phi(n) e^{-\mu n^2} = \sqrt{\frac{%
\mu^{\prime}}{\pi}} \sum\limits_{m=0}^{M-1} \phi(m)  \Theta_3(\pi
m/M|\mu^{\prime}),
\]
\[
\sum\limits_{n=-\infty}^{\infty} (-1)^n \phi(n) e^{-\mu n^2} = \sqrt{\frac{%
\mu^{\prime}}{\pi}} \sum\limits_{m=0}^{M-1} (-1)^m \phi(m)  \Theta_2(\pi
m/M|\mu^{\prime}),
\]
\[
\sum\limits_{n=-\infty}^{\infty} \phi(n) e^{-\mu (n-M/2)^2} = \sqrt{\frac{%
\mu^{\prime}}{\pi}} \sum\limits_{m=0}^{M-1} \phi(m)  \Theta_4(\pi
m/M|\mu^{\prime}),
\]
and moreover:
\[
\sum\limits_{n=-\infty}^{\infty} \phi([2n]) e^{-\mu n^2} = \sqrt{\frac{%
\mu^{\prime}}{\pi}} \sum\limits_{m=0}^{M-1} \phi(m)  \Theta_3\left(\left.
\frac{\pi}{M}\left[\frac{ m}{2}\right]\right|\mu^{\prime}\right)
\]
\[
\sum\limits_{n=-\infty}^{\infty} \phi([2n]) e^{-\mu (n-M/2)^2} = \sqrt{\frac{%
\mu^{\prime}}{\pi}} \sum\limits_{m=0}^{M-1} \phi(m)  \Theta_4\left(\left.
\frac{\pi}{M}\left[\frac{ m}{2}\right]\right|\mu^{\prime}\right).
\]
Therefore, for any periodic function of two variables, $\phi(m+M,n)=%
\phi(m,n+M)=\phi(m,n)$, one can prove,
\[
\sum_{m,n=0}^{M-1} \phi(m,n) F(m,n)=  \sum_{m,n=-\infty}^{\infty} \phi(m,n)
f(m,n) ,
\]
with
\begin{equation}  \label{ffun}
f(m,n)=\sqrt{\frac{\pi}{2\mu}}\,(-1)^{mn} \exp\left( -\frac{\mu^{\prime}}{2}%
\,m^2 -\frac{\mu }{2}\,n^2 \right).
\end{equation}
From here it follows that $F(m,n)=F(n,m)$ (provided that
$\mu=\mu'$, {\it i.e.} the coordinate and momentum have the same
rights.

\section{Conclusions}

The principal results of this paper are the relations (\ref{relQW}),(\ref
{Ffun}),(\ref{ffun}) between the discrete Q-function (time-dependent
spectrum) and the discrete Wigner function. As in the continuous case, the
Q-function can be produced by smoothing the W-function by integrating it
with the discrete Gaussian function.

S.C. is grateful to the Department of Computer Science, University of
Houston, Texas for hospitality.

\section*{Appendix A: Theta functions}

Jacobi Theta functions are the integral functions defined as follows \cite
{WhWat}:
\begin{eqnarray*}
\Theta_3(z|\mu) \!\!&=& \!\! \sum_{n=-\infty}^{\infty} \exp \left\{-i2 n z
-\mu n^2 \right\} = \sqrt{\frac{\pi}{\mu}}\sum_{k=-\infty}^{\infty} \exp
\left\{- \frac{(z-\pi k)^2}{\mu} \right\}, \\
\Theta_4(z|\mu)\!\! &=&\!\! \sum_{n=-\infty}^{\infty} (-1)^n \exp \left\{-i2
n z -\mu n^2 \right\} = \sqrt{\frac{\pi}{\mu}} \sum_{k=-\infty}^{\infty}
\exp \left\{ -\frac{\left(z-\pi (k+1/2)\right)^2}{\mu} \right\} , \\
\Theta_2(z|\mu)\!\! &=&\!\! \sum_{n=-\infty}^{\infty} \exp \left\{-i2
\left(n +\frac 12\right) z -\mu \left( n\!+\!\frac 12\right)^2 \right\} =
\sqrt{\frac{\pi}{\mu}}\sum_{k=-\infty}^{\infty} (-1)^k \exp \left\{- \frac{%
(z-\pi k)^2}{\mu} \right \} , \\
\Theta_1(z|\mu)\!\! &=&\!\! \sum_{n=-\infty}^{\infty} (-1)^n \exp \left\{-i2
\left(n +\frac 12\right) z -\mu \left( n+\frac 12\right)^2 \right\} \\
&=& \sqrt{\frac{\pi}{\mu}}\sum_{k=-\infty}^{\infty} (-1)^k \exp \left\{-
\frac{(z-\pi (k\!+\!1/2))^2}{\mu} \right \} .
\end{eqnarray*}
Here $i\mu$ is quasiperiod, Re $\mu>0.$ Note, that in \cite{WhWat} Theta
functions are written as $\Theta_k(z,q)$, where $r=e^{-\mu}$. $\Theta_{1,2,4}
$ are shifts of $\Theta_3:$
\begin{eqnarray*}
\Theta_4(z|\mu) &=& \Theta_3(z+\pi/2|\mu), \quad \Theta_2(z|\mu) = -ie^{i
z-\mu/4} \Theta_3(z+i\mu/2|\mu), \\
\Theta_1(z|\mu) &=& \Theta_2(z-\pi/2|\mu) = -ie^{i z-\mu/4}
\Theta_3(z+\pi/2+i\mu/2|\mu) .
\end{eqnarray*}
Note, that
\[
\Theta_{1,2}(z+\pi|\mu) = - \Theta_{1,2}(z|\mu) , \quad
\Theta_{3,4}(z+\pi|\mu) = \Theta_{3,4}(z|\mu),
\]
and
\[
\Theta_{1,4}(z+i\mu|\mu) = - g\Theta_{1,4}(z|\mu) , \quad
\Theta_{2,3}(z+i\mu|\mu) = g\Theta_{2,3}(z|\mu),
\]
where $g=e^{- i2 z +\mu}$. Finally,
\[
\Theta_{1}(-z|\mu) = - \Theta_{1}(z|\mu) , \quad  \Theta_{2,3,4}(-z|\mu) =
\Theta_{2,3,4}(z|\mu).
\]


\end{document}